\documentclass[12pt]{article}
\usepackage{amsmath,amssymb,amsthm,amscd,amsfonts,indentfirst,color,setspace,bbold,slashed}
\usepackage[body={15cm, 23cm, centered}]{geometry}
\usepackage[debug,pageanchor=false]{hyperref}
\hypersetup{colorlinks=true,linktocpage,breaklinks,
 	urlcolor=blue,
 	linkcolor=red,
 	citecolor=blue
 }

\numberwithin{equation}{section}

\def\be{\begin{equation}}
\def\ee{\end{equation}}
\def\ben{\begin{equation*}}
\def\een{\end{equation*}}
\def\ba{\begin{array}}
\def\ea{\end{array}}
\def\bn{\begin{aligned}}
\def\en{\end{aligned}}
\def\bnn{\begin{eqnarray*}}
\def\enn{\end{eqnarray*}}
\def\bsub{\begin{subequations}}
\def\esub{\end{subequations}}

\def\p{{\partial}}

\def\a{{\alpha}}
\def\b{{\beta}}

\def\G{{\Gamma}}
\def\g{{\gamma}}
\def\d{{\delta}}
\def\e{{\epsilon}}
\def\ve{{\varepsilon}}

\def\z{{\zeta}}
\def\h{{\eta}}
\def\th{{\theta}}

\def\l{{\lambda}}

\def\m{{\mu}}
\def\n{{\nu}}

\def\r{{\rho}}
\def\s{{\sigma}}
\def\S{{\Sigma}}

\def\f{{\phi}}

\def\c{{\chi}}
\def\ps{{\psi}}
\def\w{{\omega}}

\def\u{\underline}

\def\7{AdS_7 \times M_3}
\def\5{AdS_5 \times M_5}

\begin{document}

\begin{titlepage}
	
	\title{
		\vskip-40pt
		\begin{flushright}
			%{\small ICTP-SAIFR/2013-012}\\
			~\\
			~\\
		\end{flushright}
		{\bf Fermionic T-duality in massive type IIA supergravity on $AdS_{10-k} \times M_k$}
		~\\
		~\\
		\author{Ilya~Bakhmatov\footnote{\tt ivbahmatov@kpfu.ru}
			~\\
			~\\
			{\it Kazan Federal University, Institute of Physics}\\
			{\it General Relativity Department}\\
			{\it Kremlevskaya 16a, 420111, Kazan, Russia}
			~\\
			\date{}
		}}
		
		\maketitle
		
		\begin{abstract}
			\noindent
			Fermionic T-duality transformation is studied for the $\mathcal{N}=1$ supersymmetric solutions of massive type IIA supergravity with the metric $AdS_{10-k} \times M_k$ for $k=3$ and $5$. We derive the Killing spinors of these backgrounds and use them as an input for the fermionic T-duality transformation. The resulting dual solutions form a large family of supersymmetric deformations of the original solutions by complex valued RR fluxes. We observe that the Romans mass parameter does not change under fermionic T-duaity, and prove its invariance in the $k=3$ case.
		\end{abstract}
		~\\
		~\\
		
		\thispagestyle{empty}
	\end{titlepage}

%\begin{titlepage}
%		
%\vfill
%	
%\begin{center}
%\baselineskip=16pt
%{\Large \bf  Fermionic T-duality\\ in massive type IIA supergravity on $AdS_{10-k} \times M_k$}
%\vskip 2cm
%Ilya Bakhmatov${}^\dagger$\footnote{\tt ivbahmatov@kpfu.ru}
%\vskip .6cm
%\begin{small}
%{\it ${}^\dagger$Kazan Federal University, Institute of Physics \\
%General Relativity Department\\
%Kremlevskaya 16a, 420111, Kazan, Russia} 
%\end{small}
%\end{center}
%	
%\vfill 
%\begin{center} 
%\textbf{Abstract}
%\end{center} 
%\begin{quote}
%Fermionic T-duality transformation is studied for the $\mathcal{N}=1$ supersymmetric solutions of massive type IIA supergravity of the form $AdS_{10-k} \times M_k$ for $k=3$ and $5$. We give a detailed derivation of the Killing spinors of these backgrounds and comment on various properties of the dual solutions. We observe that the Romans mass parameter does not change under fermionic T-duaity, and prove its invariance in the $k=3$ case.
%\end{quote} 
%\vfill
%\setcounter{footnote}{0}
%\end{titlepage}

\clearpage
\setcounter{page}{2}

\tableofcontents

\section{Introduction}
\label{intro}

Families of new solutions in type II supergravity of the form $AdS_{10-k} \times M_{k}$ were found in \cite{Apruzzi:2013yva, Apruzzi:2014qva, Apruzzi:2015zna} for $k = 3,4,5$. For $k=3$ and $5$ the solutions belong to the massive type IIA supergravity~\cite{Romans:1985tz}, while for $k=4$ they solve type IIB field equations. In the $AdS_7 \times M_3$ solutions the internal manifold $M_3$ is topologically a sphere. The requirement of unbroken supersymmetry was demonstrated to fix $M_3$ to be a fibration of $\mathbb{S}^2$ over the interval. The background fields of these solutions were given by the system of first order differential equations, which the authors of~\cite{Apruzzi:2013yva} were solving numerically. An infinite family of solutions was obtained, which have embedded D6/D8 brane systems. The holographic interpretation of these theories was investigated in \cite{Gaiotto:2014lca,DelZotto:2014hpa}. 

Analytic solutions to these equations were found later, together with a map that relates them to the $AdS_5$ and $AdS_4$ solutions in massive type IIA \cite{Apruzzi:2015zna,Rota:2015aoa}. The $AdS_5$ solutions that we will be concerned with in the present paper are geometrically $AdS_5 \times \S_2 \times M_3'$, where $\S_2$ is a Riemann surface of genus $g\geq2$, and $M_3'$ is a three-manifold related in a certain way to $M_3$. A recent review of these and related developments can be found in~\cite{Apruzzi:2015wna}.

The aim of this paper is to study the effect of fermionic T-duality on the $AdS_{10-k} \times M_{k}$ solutions for $k=3$ and $k=5$. It is a well known fact that the transformation rules of the background fields under ordinary T-duality (known also as the Buscher rules~\cite{Buscher:1985kb,Buscher:1987sk,Buscher:1987qj}) can be represented in a way which manifests the role of the Killing vector as the T-duality transformation parameter~\cite{Alvarez:1993qi}. Fermionic T-duality is a more recent development~\cite{Berkovits:2008ic,Beisert:2008iq}, which generalises T-duality to the case of fermionic isometries in superspace. The role of a Killing vector is played by a Killing spinor, which parameterises an unbroken supersymmetry of the initial background. The fermionic T-dual background can then be constructed according to the fermionic Buscher rules, which depend explicitly on the Killing spinor. The key difference from the ordinary T-duality rules is that the metric and the NSNS 2-form field $b$ do not change, whereas the RR fluxes are transformed in a certain way that depends on the Killing spinors of the original background. Fermionic T-duality plays a key role in the self-duality of various solutions of maximal $d=10$ supergravity that are important from the AdS/CFT correspondence point of view. Self-duality under a set of combined bosonic and fermionic T-dualities has been observed for $AdS_5 \times \mathbb{S}^5$~\cite{Berkovits:2008ic}, for pp-wave spacetimes~\cite{Bakhmatov:2011aa}, for $AdS_3\times \mathbb{S}^3\times\mathbb{T}^4$~\cite{OColgain:2012ca}, and most recently for $AdS_d \times \mathbb{S}^d \times \mathbb{T}^{10-2d}$ and $AdS_d \times \mathbb{S}^d \times \mathbb{S}^d \times \mathbb{T}^{10-3d}$ ($d=2,3$)~\cite{Abbott:2015mla,Abbott:2015ava}. 

In order to construct fermionic T-duals of the solutions of~\cite{Apruzzi:2013yva,Apruzzi:2015zna} we study the unbroken supersymmetries of these backgrounds and solve the Killing spinor equations in full generality. Note that concise expressions for the $AdS_7 \times M_3$ Killing spinors have appeared in \cite{Rota:2015aoa,Passias:2015gya}, while the Killing spinor structure of the $AdS_6 \times M_4$ solutions has been studied in detail in \cite{Kim:2015hya}. Fermionic T-duality preserves supersymmetry and the metric of the solution~\cite{Berkovits:2008ic}, hence using the Killing spinors we are able to generate new supersymmetric solutions with the same metric $AdS_{10-k} \times M_k$, $k=3,5$. The new solutions presented here are essentially deformations of the original solutions by complex valued RR fluxes, akin to the deformations of the D-brane solutions found earlier in~\cite{Bakhmatov:2009be}.

The behaviour of massive type IIA supergravity solutions under fermionic T-duality is an open question since the early work~\cite{Godazgar:2010ph}, where a fermionic duality symmetry of type II supergravity action has been found that includes fermionic T-duality as a special case. While formally applicable to both ordinary and massive type IIA supergravity, the analysis of~\cite{Godazgar:2010ph} assumed vanishing mass parameter $m$ before the duality transformation, and resulted in keeping $m$ zero after the duality as well. It was later reported~\cite{OColgain:2012si}, that the extension of the transformation of~\cite{Godazgar:2010ph} to a nonzero Romans mass, when applied to characteristic solutions of massive type IIA, such as D8-branes and the warped product $AdS_6 \times \mathbb{S}^4$~\cite{Brandhuber:1999np}, yields no change in the mass parameter, and that the entire transformation is trivial in that case. In the current study we will be using the original fermionic T-duality formalism developed in \cite{Berkovits:2008ic}. We will give a proof that the Romans mass of the $AdS_7 \times M_3$ solutions of \cite{Apruzzi:2013yva} never changes under fermionic T-duality. In the less tractable case of the $AdS_5 \times M_5$ solutions of \cite{Apruzzi:2015zna} evidence will be given that the Romans mass does not change, although this will not be proved rigorously.

The rest of this article is organised as follows. We begin in the section~\ref{ads7} by studying the simpler case of the $AdS_7\times M_3$ solutions. Then in the section~\ref{ads5} the $AdS_5\times M_5$ solutions are considered. In both cases we briefly review the solutions, then formulate the Killing spinor equations and solve them. Section~\ref{ferm} presents the fermionic T-duals after a concise review of the fermionic Buscher rules. We briefly discuss the results in the final section~\ref{conc}. Our notation and conventions are summarised in the appendix~\ref{gam}.

\section{$AdS_7 \times M_3$ solution}
\label{ads7}

The $AdS_7 \times M_3$ background of~\cite{Apruzzi:2013yva} is an $\mathcal{N}=1$ supersymmetric solution in massive type IIA supergravity. The metric is given by
\be\label{metric-7x3}
\begin{gathered}
	ds^2 = e^{2A(r)} ds^2_{AdS_7} + ds^2_{M_3},\\
	ds^2_{AdS_7} = \r^2\left[ -(dx^0)^2 + (dx^1)^2 + \ldots + (dx^5)^2\right]  + \frac{d\r^2}{\r^2},\\
	ds^2_{M_3} = dr^2 + \frac{e^{2A(r)}}{16} (1-x(r)^2) \left( d\b^2 + \sin^2\b\,d\th^2 \right).
\end{gathered}
\ee
The ten spacetime coordinates $x^\m$ are split into the $AdS_7$ coordinates $(x^0,\ldots,x^5,\r)$ and the three coordinates $(r,\b,\th)$ on the internal manifold. $M_3$ is an $\mathbb{S}^2$ fibration over an interval that is parameterised by the coordinate $r$. The $\mathbb{S}^2$ fibre shrinks at the ends of the interval, so that $M_3$ is topologically a 3-sphere. The warping function $A(r)$, as well as the dilaton $\f(r)$, and the parameter $x(r)$ of the internal metric, depend on $r$ only. The function $x(r)$ is related to the volume of the $\mathbb{S}^2$ fibre. These three functions are defined by the following system of differential equations:		
\be\label{back-der-7x3}\bn
\f'(r) &= \frac14 \frac{e^{-A}}{\sqrt{1-x^2}}\, \left( 12x + (2x^2 - 5) m e^{A+\f} \right),\\
x'(r) &= -\frac12 e^{-A} \sqrt{1-x^2} (4+x m e^{A+\f}),\\
A'(r) &= \frac14 \frac{e^{-A}}{\sqrt{1-x^2}}\, (4x - m e^{A+\f}),
\en\ee
where $m$ is a constant mass parameter of Romans supergravity. The authors of~\cite{Apruzzi:2013yva} study numerical solutions to this system. Later an analytic solution of these equations has been found in~\cite{Apruzzi:2015zna}; it describes backgrounds with D6 or D8 branes. For our purposes the equations~\eqref{back-der-7x3} will be enough, and we will not spell out the details of the explicit solutions.

The metric is diagonal, and we can choose the vielbein in the simple form:
\be\label{vielbein}
\begin{gathered}
e_{\underline 0}^0 = \ldots = e_{\underline 5}^5 = \r e^A,\qquad e_{\underline \r}^6 = \frac{e^A}{\r},\\ 
e_{\underline r}^7 = 1,\qquad e_{\underline \b}^8 = \frac{e^A}{4}\sqrt{1-x^2},\qquad e_{\underline \th}^9 = \frac{e^A}{4}\sqrt{1-x^2}\sin\b.
\end{gathered}
\ee
We underline the world indices and assume that $dx^{\u \m} = (dx^0,\dots, dx^5,d\r,dr,d\b,d\th)$. Then the vielbein $e^a = e_{\u \m}^a \,dx^{\u \m}$ corresponds to the metric~\eqref{metric-7x3}, $ds^2 = \h_{ab}e^a e^b$. With this choice of the vielbein the nonvanishing components of internal spin connection are:
\be
\label{spin-conn-7x3}
\begin{gathered}
\w_{\u \b,78} = \frac{e^A}{4\sqrt{1-x^2}}\, \left(xx' - A' (1-x^2)\right),\quad \w_{\u \th,79} = \frac{e^A \sin\b}{4 \sqrt{1-x^2}} \left(xx' - A'(1-x^2)\right),\\ \w_{\u\th,89} = -\cos\b.
\end{gathered}
\ee

The other nonvanishing fields of the supergravity background are the RR 2-form and the NSNS $H$ flux:
\be\label{flux-7x3}
\begin{gathered}
	F = F_{89}\, e^8 \wedge e^9 = -\frac{e^{-\f}}{\sqrt{1-x^2}} \left( 4e^{-A} -x m e^\f \right) e^8 \wedge e^9,\\
	H = H_{789}\, e^7 \wedge e^8 \wedge e^9 =-\left( 6e^{-A} +x m e^\f \right) e^7 \wedge e^8 \wedge e^9.
\end{gathered}
\ee

\subsection{Killing spinors}
\label{killing-7x3}

Let us solve the Killing spinor equations for this background. This requires finding a spinor $\e$ such that the supersymmetry variations of the type IIA fermions vanish, $\d_\e\l =\nobreak 0 =\nobreak \d_\e\ps_\m$. Our supersymmetry and spinor conventions are summarised in the appendix~\ref{gam}. After decomposing the dilatino supersymmetry variation~\eqref{variations} with respect to the Weyl components of the Killing spinor~\eqref{epsilon-5x5}, we obtain:
\be
\d\l = \p_7\f\, {(\mathbb{1}\otimes\s^1) \ve_2 \choose (\mathbb{1}\otimes\s^1) \ve_1} + \frac{i}{2} H_{789} {-\ve_2 \choose \ve_1} + \frac{5me^\f}{4} {\ve_1 \choose \ve_2} + \frac{3ie^\f}{4} F_{89} {(\mathbb{1}\otimes\s^1) \ve_1 \choose -(\mathbb{1}\otimes\s^1) \ve_2},
\ee
where $\ve_{1,2}$ are 16-component Weyl spinors, defined in~\eqref{epsilon-5x5}. Note that $\p_7\f = \p_{\underline r} \f$~\eqref{back-der-7x3}, since $e_{\underline r}^7 = 1$, and the values of $H_{789}, F_{89}$ can be read off from~\eqref{flux-7x3}. 

Using the decomposition of $\ve_{1,2}$ in terms of the $AdS_7$ spinors $\z$ and $M_3$ spinors $\c_{1,2}$~\eqref{epsilon-5x5}, $\d\l$ may be brought to the form: 
\be
\d\l = {\z\otimes\d_1\l + \z^c \otimes \d_1\l^c \choose \z\otimes\d_2\l - \z^c \otimes \d_2\l^c },
\ee
where
\be\label{dilatino}\bn
\d_1\l = \p_7\f\,(\s^1\c_2) - \frac{i}{2} H_{789}\, \c_2 + \frac{5me^\f}{4}\, \c_1 + \frac{3ie^\f}{4} F_{89}\, (\s^1\c_1),\\
\d_2\l = \p_7\f\,(\s^1\c_1) + \frac{i}{2} H_{789}\, \c_1 + \frac{5me^\f}{4}\, \c_2 - \frac{3ie^\f}{4} F_{89}\, (\s^1\c_2).
\en\ee
Requiring that both $\d_1\l$ and $\d_2\l$ vanish imposes four linear equations on the four components of the internal Killing spinors $\c_1 = {a' \choose b'}$, $\c_2 = {a \choose b}$. The resulting system has rank two, and can be solved in a straightforward manner. For example, we can express $a', b'$ in terms of $a, b$:
\be\bn
a' =+\frac{BC+AD}{C^2-D^2} a - \frac{AC+BD}{C^2-D^2} b,\\
b' =-\frac{AC+BD}{C^2-D^2} a + \frac{BC+AD}{C^2-D^2} b,
\en\ee
where we have denoted the coefficients in~\eqref{dilatino} as:
\be
A = \p_7 \f,\qquad B = \frac{i}{2} H_{789},\qquad C = \frac{5m e^\f}{4},\qquad D = \frac{3i e^\f}{4} F_{89}.
\ee
Using the explicit values of $A,B,C$, and $D$ from~\eqref{back-der-7x3} and~\eqref{flux-7x3}, we observe that the expressions simplify considerably:
\be
\frac{AC+BD}{C^2-D^2} = -\sqrt{1-x^2},\qquad \frac{BC+AD}{C^2-D^2} = -ix,
\ee
where $x = x(r)$ was defined in~\eqref{back-der-7x3}. To summarize, the dilatino supersymmetry variation vanishes for the following values of the internal spinors:
\be\label{dil-otvet}
\c_1 = {-ixa + \sqrt{1-x^2}\,b \choose -ix b + \sqrt{1-x^2}\,a},\qquad \c_2 = {a \choose b}.
\ee
At this point $a$ and $b$ are two arbitrary functions of the coordinates $(r,\b,\th)$ on $M_3$. Their values are fixed by the Killing spinor equations that follow from $\d\ps_{i+6}=0$ for $i = 1,2,3$ ($\ps_{i+6}$ are the components of the gravitino with their vector indices belonging to the internal manifold). For the background of~\eqref{metric-7x3},~\eqref{flux-7x3} the gravitino variation~\eqref{variations} becomes:
\be\bn
\d\ps_{i+6} &= D_{i+6} {\ve_1 \choose \ve_2} + \frac{i}{4} H_{789} {(\mathbb{1} \otimes \s_i)\ve_1 \choose -(\mathbb{1} \otimes \s_i)\ve_2} + \frac{me^\f}{8} {(\mathbb{1} \otimes \s_i)\ve_2 \choose (\mathbb{1} \otimes \s_i)\ve_1} \\&+ \frac{ie^\f}{8} F_{89} {-(\mathbb{1} \otimes \s_1\s_i)\ve_2 \choose (\mathbb{1} \otimes \s_1\s_i)\ve_1}.
\en\ee
Under the $7+3$ split of~\eqref{epsilon-5x5} the $AdS_7$ spinor $\z$ factors out,
\be
\d\ps_{i+6} = {\z\otimes\d_1\ps_{i+6} + \z^c \otimes \d_1\ps_{i+6}^c \choose \z\otimes\d_2\ps_{i+6} - \z^c \otimes \d_2\ps_{i+6}^c },
\ee
where
\be\label{gravitino}\bn
\d_1\ps_{i+6} = D_{i+6}\, \c_1 + \frac{i}{4} H_{789}\, (\s_i \c_1) + \frac{me^\f}{8}\, (\s_i \c_2) - \frac{ie^\f}{8} F_{89}\, (\s_1 \s_i \c_2),\\
\d_2\ps_{i+6} = D_{i+6}\, \c_2 - \frac{i}{4} H_{789}\, (\s_i \c_2) + \frac{me^\f}{8}\, (\s_i \c_1) + \frac{ie^\f}{8} F_{89}\, (\s_1 \s_i \c_1).
\en\ee
The covariant derivative acting on a spinor is $D_\m = \p_\m + \frac12 \slashed \w_\m$. Note that since all the mixed components $\w_{i+6,mn}$ of the spin connection vanish~\eqref{spin-conn-7x3}, the derivative of $\c_{1,2}$ simplifies to
\be
D_{i+6}\, \c = \p_{i+6}\, \c + \frac14 \w_{i+6,j+6,k+6}\s^{jk}.
\ee
Using the values of $H_{789}$, $F_{89}$ from~\eqref{flux-7x3} and the internal spinors $\c_{1,2}$ found in~\eqref{dil-otvet}, after some algebra one can represent the equations~\eqref{gravitino} as
\be
\begin{gathered} 
	\p_{\u r}\, a -\frac{A'(r)}{2}\, a - \frac{i}{2} \frac{x'(r)}{\sqrt{1-x^2}}\, b = 0,\\
	\p_{\u r}\, b -\frac{A'(r)}{2}\, b - \frac{i}{2} \frac{x'(r)}{\sqrt{1-x^2}}\, a = 0,\\
	\p_{\u \b}\, a - \frac{i}{2}\, x(r)\, a + \frac12 \sqrt{1-x^2}\, b = 0,\\
	\p_{\u \b}\, b + \frac{i}{2}\, x(r)\, b - \frac12 \sqrt{1-x^2}\, a = 0,\\
	\p_{\u \th}\, a + \frac{i}{2}\,\sin\b\,\sqrt{1-x^2}\,a + \frac12 \left( x\sin\b - i\cos\b \right)\,b = 0,\\
	\p_{\u \th}\, b - \frac{i}{2}\,\sin\b\,\sqrt{1-x^2}\,b - \frac12 \left( x\sin\b + i\cos\b \right)\,a = 0.
\end{gathered}
\ee
Despite the presence of arbitrary functions $A(r), x(r)$, this system can be solved exactly:
\be\label{chi-otvet}
\begin{gathered}
	a=e^\frac{A}{2} \left[ c_1 e^{i\frac{\th}{2}} \left( \cos\frac{\b}{2}\, e^{i\varphi} - \sin\frac{\b}{2}\, e^{-i\varphi} \right) + c_2 e^{-i\frac{\th}{2}} \left( \sin\frac{\b}{2}\, e^{i\varphi} + \cos\frac{\b}{2}\, e^{-i\varphi} \right) \right],\\
	b=e^\frac{A}{2} \left[ c_1 e^{i\frac{\th}{2}} \left( \cos\frac{\b}{2}\, e^{i\varphi} + \sin\frac{\b}{2}\, e^{-i\varphi} \right) + c_2 e^{-i\frac{\th}{2}} \left( \sin\frac{\b}{2}\, e^{i\varphi} - \cos\frac{\b}{2}\, e^{-i\varphi} \right) \right],
\end{gathered}
\ee
where $\varphi$ is a new internal variable related to $r$ by $\sin 2 \varphi = x(r)$. %,\qquad e^{2 i\varphi} = ix(r) + \sqrt{1-x(r)^2}. 
The solution is parameterised by the two constants $c_{1,2}$. Together with~\eqref{dil-otvet}, this completely determines the internal part of the Killing spinor, $\c_{1,2}$. 

Simplifying the $AdS_7$ part of the gravitino Killing spinor equation $\d\ps_m = 0$, we get the standard expression for the $AdS_7$ Killing spinor, as briefly reviewed in the appendix~\ref{ads}:
\be\label{zeta}
\z = \r^{1/2} \z_-^0 + \left( \r^{-1/2} - \r^{1/2} x^m \a_m \right) \z_+^0, \qquad m\in\{0,\ldots,5\}.
\ee
Here $\z^0$ is an arbitrary 8 component complex spinor parameter, and $\z^0 = \z^0_+ + \z^0_-$ is its decomposition into eigenvectors of $\a_6$, which is the gamma matrix corresponding to the $AdS_7$ radial direction $\r$. The complete Killing spinors are then given by~\eqref{epsilon-5x5}:
\be\label{otvet}
\e = {\z \otimes \c_1 + \z^c \otimes \c_1^c \choose \z \otimes \c_2 - \z^c \otimes \c_2^c}.
\ee
It is easy to see that there is a total of 16 Killing spinors $\e_a$, $a=1,\ldots,16$, which means that the solution of~\cite{Apruzzi:2013yva} preserves half of the maximal supersymmetry. To check this, note that we can represent the eigenvectors of $\a_6$~\eqref{small-gamma7x3} as 
\be\label{zeta-numbers}\bn
\z^0_+ &= \left[ -i(\z^0_9+i\z^0_{10}), i(\z^0_{11}+i\z^0_{12}), i(\z^0_{13}+i\z^0_{14}), -i(\z^0_{15}+i\z^0_{16}),\right. \\ &\left. \qquad\z^0_{15}+i\z^0_{16}, \z^0_{13}+i\z^0_{14}, \z^0_{11}+i\z^0_{12}, \z^0_9+i\z^0_{10} \right],\\
\z^0_- &= \left[  i(\z^0_1+i\z^0_2),-i(\z^0_3+i\z^0_4), -i(\z^0_5+i\z^0_6),  i(\z^0_7+i\z^0_8),\right. \\ &\left. \qquad\z^0_7+i\z^0_8, \z^0_5+i\z^0_6, \z^0_3+i\z^0_4, \z^0_1+i\z^0_2 \right],
\en\ee
in terms of 16 real components of $\z^0 = (\z^0_1,\ldots,\z^0_{16})$. Taking $c_1=1, c_2=0$ in~\eqref{chi-otvet} and setting $\z^0_a = \d_{ab}$ for some $b \in \{1,\ldots,16\}$ results in the Killing spinor that we will call $\e_b$. Explicit computation then shows, that taking other values of the parameters $c_1, c_2$ in~\eqref{chi-otvet} results in the same set of Killing spinors up to relabeling.

Note that the numbering convention of~\eqref{zeta-numbers} implies that $\e_1,\ldots,\e_8$ are the Poincar\'e Killing spinors that only depend on the radial $AdS_7$ coordinate $\r$, while $\e_9,\ldots,\e_{16}$ are the superconformal Killing spinors that depend on all $AdS_7$ coordinates, cf.~\eqref{zeta}.

\section{$AdS_5 \times M_5$ solutions}
\label{ads5}

The $AdS_5 \times M_5$ solution of~\cite{Apruzzi:2015zna} has the metric given by
\be\label{metric-5x5}
\begin{gathered}
ds^2 = e^{2A(r)} \left( ds^2_{AdS_5} + ds^2_{\S_g} \right) + ds^2_{M_3},\\
ds^2_{AdS_5} = \r^2\left[ -(dx^0)^2 + (dx^1)^2 + (dx^2)^2 + (dx^3)^2\right]  + \frac{d\r^2}{\r^2},\\
ds^2_{\S_g} = \frac{1}{-k\,x_2^2} \left( dx_1^2 + dx_2^2 \right),\\
ds^2_{M_3} = dr^2 + \frac{e^{2A(r)}}{9} \left(1-x(r)^2\right) \left( d\th^2 + \sin^2\th\,D\ps^2 \right).
\end{gathered}
\ee
$\S_g$ is a Riemann surface of Gaussian curvature $k=-3$ and genus $g\geq2$. Its metric is written in terms of the coordinate $z=x_1+ix_2$ of the complex upper half-plane. Coordinates on the Riemann surface $x_1,x_2$ are not to be confused with the $AdS_5$ coordinates $x^0,\ldots,x^3,\r$ (we will never raise or lower indices of coordinates).

The $M_3$ subspace of the internal manifold is fibered over the Riemann surface $\S_g$, which is reflected by the long derivative $D\ps = d\ps + \r$ appearing in the metric $ds^2_{M_3}$. Here the 1-form $\r$ is defined by $\r = \r_1(x_1,x_2) dx_1 + \r_2 (x_1,x_2) dx_2$. This 1-form is subject to the constraint
\be
d_2 \r = k\, \mathrm{vol}_{\S_g},
\ee
which for the above metric takes the form 
\be\label{rho-constraint}
\p_1 \r_2 - \p_2 \r_1 = -(x_2)^{-2}.
\ee

The warping function $A(r)$, as well as the parameter $x(r)$ of the internal metric, and the dilaton $\f(r)$, only depend on the coordinate $r$, which runs over an interval. All of these are defined by the following system of ODEs:		
\be\label{back-der-5x5}\bn
\f'(r) &= \frac14 \frac{e^{-A}}{\sqrt{1-x^2}} \left( 11x -2x^3 + (2x^2 - 5) m e^{A+\f} \right),\\
x'(r) &= -\frac12 e^{-A} \sqrt{1-x^2} \,(4 -x^2 + m x e^{A+\f}),\\
A'(r) &= \frac14 \frac{e^{-A}}{\sqrt{1-x^2}}\, (3x - m e^{A+\f}),
\en\ee
where $m$ is a constant mass parameter of Romans supergravity. Note there are slight differences in the coefficients as compared to the $AdS_7 \times M_3$ case~\eqref{back-der-7x3}. As was already mentioned there, we will not need the explicit form of the solutions to this system of equations.

We can choose the following simple vielbein:
\be\label{vielbein-5x5}
\begin{gathered}
e_{\u 0}^0 = \ldots = e_{\u 3}^3 = \r\, e^A,\qquad e_{\u \r}^4 = \r^{-1}e^A, \qquad e_{\u x_1}^5 = e_{\u x_2}^6 = e^A(-k)^{-1/2} (x_2)^{-1},\\
\qquad e_{\u r}^7 = 1,\qquad e_{\u \th}^8 = \frac{e^A}{3}\sqrt{1-x^2},\qquad e_{\u \ps}^9 = \frac{e^A}{3}\sqrt{1-x^2}\sin\th,\\ \qquad e_{\u x_1}^9 = \frac{e^A}{3}\sqrt{1-x^2}\, \r_1 \sin\th,\qquad e_{\u x_2}^9 = \frac{e^A}{3}\sqrt{1-x^2}\,\r_2\sin\th.
\end{gathered}
\ee
We underline the world indices and assume that $dx^{\u \m} = (dx^0,\dots, dx^3, d\r, dx_1, dx_2, \linebreak dr, d\th, d\ps)$. Then the vielbein $e^a = e_{\u \m}^a \,dx^{\u \m}$ corresponds to the metric~\eqref{metric-5x5}, $ds^2 = \h_{ab}e^a e^b$. With this choice of the vielbein the nonvanishing components of internal spin connection are:
\be
\label{spin-conn-5x5}
\begin{gathered}
\w_{\u x_1,56} = -\frac{1}{x_2} + \frac{\r_1}{6} \left(1-x^2\right) \sin^2 \th, \quad \w_{\u x_1,57} = \frac{A' e^A}{\sqrt{3}\, x_2}, \quad \w_{\u x_1,69} = -\frac{\sin\th \sqrt{1-x^2}}{2\sqrt{3}\, x_2},\\ 
\quad \w_{\u x_1,79} = \frac{\r_1\, e^A \sin\th}{3 \sqrt{1-x^2}} \left(xx' - A'(1-x^2)\right), \quad \w_{\u x_1,89} = -\r_1 \cos\th;\\
\w_{\u x_2,56} = \frac{\r_2}{6} \left(1-x^2\right) \sin^2 \th, \quad \w_{\u x_2,67} = \frac{A' e^A}{\sqrt{3}\, x_2}, \quad \w_{\u x_2,59} = \frac{\sin\th \sqrt{1-x^2}}{2\sqrt{3}\, x_2},\\ 
\quad \w_{\u x_2,79} = \frac{\r_2\, e^A \sin\th}{3 \sqrt{1-x^2}} \left(xx' - A'(1-x^2)\right), \quad \w_{\u x_2,89} = -\r_2 \cos\th;\\
\w_{\u r} \equiv 0;\quad \w_{\u \th,78} = \frac{e^A}{3\sqrt{1-x^2}}\, \left(xx' - A' (1-x^2)\right); \quad
\w_{\u \ps,56} = \frac16 (1-x^2) \sin^2\th,\\ \quad \w_{\u \ps,79} = \frac{e^A \sin\th}{3 \sqrt{1-x^2}} \left(xx' - A'(1-x^2)\right),\quad \w_{\u\ps,89} = -\cos\th.
\end{gathered}
\ee

The nonvanishing fluxes of the solution are
\be\label{flux-5x5}
\begin{gathered}
F_{(2)} = -e^{-A-\f} \sqrt{1-x^2}\, \cos\th\, e^5 \wedge e^6 - e^{-A-\f} \frac{\left(3 -m x e^{A+\f} \right)}{\sqrt{1-x^2}}\, e^8 \wedge e^9,\\
F_{(4)} = -e^{-A-\f}\, e^5 \wedge e^6 \wedge \left( e^7\sin\th + e^8 x \cos\th \right) \wedge e^9,\\
H = -e^{-A}\, e^5\wedge e^6\wedge \left( e^7 \cos\th - e^8 x\sin\th \right) + e^{-A} \left( x^2 - 5 - mx e^{A+\f} \right) e^7\wedge e^8\wedge e^9.
\end{gathered}
\ee

\subsection{Killing spinors}
\label{killing-5x5}

Let us construct the Killing spinors for the above background. As in the section~\ref{ads7}, we start with the variation of the dilatino~\eqref{variations}. Plugging in the values of the fluxes~\eqref{flux-5x5} and using the $5+5$ decomposition of the 16-component Weyl spinors $\ve_{1,2}$~\eqref{epsilon-5x5}, we find that the dilatino variation can be written:
\be
\d\l = {\z\otimes\d_1\l + \z^c \otimes \d_1\l^c \choose \z\otimes\d_2\l - \z^c \otimes \d_2\l^c },
\ee
where
\be\label{dil-ads5}\bn
\d_1\l &= A_1\,\b^3\c_2 + A_2\, \c_1 +A_3\,\b^4\c_1 - A_4\,\b^3\c_1 - A_5\,\b^4\b^5 \c_2 + A_6\,\b^3\b^5 \c_2 - A_7\,\b^1\b^2 \c_2\\ &+ A_8\, \b^1\b^2\c_1 + A_9\, \b^4\b^5\c_1,\\
\d_2\l &= A_1\,\b^3\c_1 + A_2\, \c_2 +A_3\,\b^4\c_2 - A_4\,\b^3\c_2 + A_5\,\b^4\b^5 \c_1 - A_6\,\b^3\b^5 \c_1 + A_7\,\b^1\b^2 \c_1\\ &- A_8\, \b^1\b^2\c_2 - A_9\, \b^4\b^5\c_2,
\en\ee
and the coefficients are given by:
\be
\begin{gathered}
A_1 = \p_r\f,\quad A_2 = \frac{5me^\f}{4},\quad A_3 = \frac{e^\f}{4} F_{5679},\quad A_4 = \frac{e^\f}{4} F_{5689},\\ \quad A_5 = \frac{1}{2} H_{567},\quad A_6 = \frac{1}{2} H_{568},\quad A_7 = \frac{1}{2} H_{789},\quad A_8 = \frac{3e^\f}{4} F_{56},\quad A_9 =  \frac{3e^\f}{4} F_{89}.
\end{gathered}
\ee
Their explicit values can be read off from~\eqref{flux-5x5}, \eqref{back-der-5x5}. Note that we are using a different ($5+ 5$) gamma-matrix decomposition in the present case, see appendix~\ref{gam}.

Requiring that $\d_1\l, \d_2\l$ in \eqref{dil-ads5} vanish results in a system of eight linear homogeneous equations for the eight unknown components of $\c_1 = (f_1,\ldots,f_4)$ and $\c_2 = (f_5,\ldots,f_8)$. Determinant of this system vanishes due to a rather non-trivial relationship between the coefficients:
\be\bn
&\left(A_1^2-A_2^2+A_3^2-A_4^2+A_5^2-A_6^2+A_7^2-A_8^2+A_9^2\right)^2\\
&-4 \left(A_1^2
A_9^2\right.+A_2^2 A_4^2+A_2^2
A_6^2-A_2^2 A_9^2-A_3^2
A_4^2+A_3^2 A_7^2-A_3^2
A_8^2\\ &\left.+A_3^2 A_9^2-A_4^2
A_9^2-A_5^2 A_6^2+A_5^2
A_7^2+A_5^2 A_9^2+A_6^2
A_8^2-A_6^2 A_9^2+A_7^2 A_9^2\right)\\
&+8(A_2 A_3 A_6 A_7-A_2 A_4 A_5
A_7+A_2 A_4 A_8 A_9+A_3 A_4
A_5 A_6+A_5 A_7 A_8 A_9)\\
&\equiv 0.
\en\ee
The rank of the corresponding matrix is six, which means that the system may be solved, for instance, for $(f_1,\ldots,f_6)$ in terms of $(f_7,f_8)$:
\be\label{dilatino-5x5}
\begin{gathered}
f_1 = \frac{f_8\cos\th - f_7}{\sin\th},\quad
f_2 = \frac{f_8 - f_7\cos\th}{\sin\th},\\
f_3 = \frac{f_7\,x\cos\th + f_8 (\sin\th-x)}{\sin\th\sqrt{1-x^2}},\quad
f_4 = \frac{f_7 (\sin\th + x) - f_8\,x\cos\th}{\sin\th\sqrt{1-x^2}},\\
f_5 = \frac{f_8 (1-x\sin\th) - f_7\cos\th}{\sin\th\sqrt{1-x^2}},\quad
f_6 = \frac{f_8\cos\th - f_7 (1+x\sin\th)}{\sin\th\sqrt{1-x^2}}.
\end{gathered}
\ee
At this point $f_7,f_8$ are two arbitrary functions of the coordinates $(x_1,x_2,r,\th,\ps)$ on $M_5$. Their values are fixed by the Killing spinor equations that follow from $\d\ps_{i+4}=0$ for $i = 1,\ldots,5$ ($\ps_{i+4}$ are the components of the gravitino with the vector index along the internal manifold). Going through the same steps as in the $AdS_7 \times M_3$ case, we obtain the following system of PDEs for the functions $f_7, f_8$:
\be\label{grav-eqns}
\begin{gathered} 
\p_{\u x_1}\, f_7 + \left(\frac{1}{x_2} + \r_1(x_1,x_2)\right) \frac{f_7\cos\th - f_8 (1-x\sin\th)}{2\sin\th\sqrt{1-x^2}} = 0,\\
\p_{\u x_1}\, f_8 - \left(\frac{1}{x_2} + \r_1(x_1,x_2)\right) \frac{f_8\cos\th - f_7 (1+x\sin\th)}{2\sin\th\sqrt{1-x^2}} = 0,\\
\p_{\u x_2}\, f_7 + \r_2(x_1,x_2) \frac{f_7\cos\th - f_8 (1-x\sin\th)}{2\sin\th\sqrt{1-x^2}} = 0,\\
\p_{\u x_2}\, f_8 - \r_2(x_1,x_2) \frac{f_8\cos\th - f_7 (1+x\sin\th)}{2\sin\th\sqrt{1-x^2}} = 0,\\
\p_{\u r}\, f_7 -\frac{A'}{2}\, f_7 + \frac{x'}{2(1-x^2)}\left(\frac{f_7 - f_8\cos\th}{\sin\th} + f_7\,x\right) = 0,\\
\p_{\u r}\, f_8 -\frac{A'}{2}\, f_8 - \frac{x'}{2(1-x^2)}\left(\frac{f_8 - f_7\cos\th}{\sin\th} - f_8\,x\right) = 0,\\
\p_{\u \th}\, f_7 - \frac{f_7\cos\th -f_8 }{2\sin\th} = 0,\qquad
\p_{\u \ps}\, f_7 + \frac{f_7\cos\th -f_8 (1-x\sin\th)}{2\sin\th\sqrt{1-x^2}} = 0,\\
\p_{\u \th}\, f_8 - \frac{f_8\cos\th -f_7 }{2\sin\th}= 0,\qquad
\p_{\u \ps}\, f_8 - \frac{f_8\cos\th -f_7 (1+x\sin\th)}{2\sin\th\sqrt{1-x^2}} = 0.
\end{gathered}
\ee
Despite explicit dependence on many arbitrary functions, this system can be solved exactly. The solution is parameterised by the two numbers $c_1, c_2$:
\be\label{grav-sol}
\begin{gathered}
f_7 = e^\frac{A}{2} \left[ c_1 e^{\frac{i}{2}(\ps+\chi)} \left(\cos\frac{\th+\varphi}{2}-i\sin\frac{\th-\varphi}{2}\right) + c_2 e^{-\frac{i}{2}(\ps+\chi)} \left(\cos\frac{\th+\varphi}{2}+i\sin\frac{\th-\varphi}{2}\right)\right],\\
f_8 = e^\frac{A}{2} \left[ c_1 e^{\frac{i}{2}(\ps+\chi)} \left(\cos\frac{\th-\varphi}{2}+i\sin\frac{\th+\varphi}{2}\right) + c_2 e^{-\frac{i}{2}(\ps+\chi)} \left(\cos\frac{\th-\varphi}{2}-i\sin\frac{\th+\varphi}{2}\right)\right],
\end{gathered}
\ee
where $\varphi$ is a new internal variable related to $r$ by $x(r)=\sin\varphi$. We have also defined
\be
\chi(x_1,x_2) = \int \r_2(x_1,x_2)dx_2.
\ee
Note the following identity, which can be obtained by integrating the constraint~\eqref{rho-constraint} with respect to $x_1$ and $x_2$:
\be
\int\r_2 dx_2 - \int\r_1 dx_1 = \frac{x_1}{x_2}.
\ee
This relationship can be employed in order to represent $\chi$ in different ways.

From the $AdS_5$ part of the gravitino Killing spinor equation $\d\ps_m = 0$ we get the standard expression for the $AdS_5$ Killing spinor, as briefly reviewed in the appendix~\ref{ads}:
\be\label{zeta1}
\z = \r^{1/2} \z_-^0 + \left( \r^{-1/2} - \r^{1/2} x^m \a_m \right) \z_+^0, \qquad m\in\{0,\ldots,3\}.
\ee
Here $\z^0$ is an arbitrary 4 component complex spinor parameter, and $\z^0 = \z^0_+ + \z^0_-$ is its decomposition into the eigenspinors of $\a_4$, which is the gamma matrix corresponding to the $AdS_5$ radial direction $\r$. The complete Killing spinors are then given by~\eqref{epsilon-5x5}:
\be\label{otvet-5x5}
\e = {\z \otimes \c_1 + \z^c \otimes \c_1^c \choose \z \otimes \c_2 - \z^c \otimes \c_2^c}.
\ee
It is easy to see that there is a total of 16 Killing spinors $\e_a$, $a=1,\ldots,16$, which means that the solution of~\cite{Apruzzi:2015zna} preserves half of the maximal supersymmetry. To check this, note that we can represent the eigenvectors of $\a_4$~\eqref{small-gamma5x5} in the form 
\be\label{zeta-numbers1}\bn
\z^0_+ &= \left[ \z^0_5+i\z^0_6, \z^0_7+i\z^0_8, \z^0_7+i\z^0_8, -(\z^0_5+i\z^0_6) \right],\\
\z^0_- &= \left[ \z^0_1+i\z^0_2, \z^0_3+i\z^0_4, -(\z^0_3+i\z^0_4), \z^0_1+i\z^0_2 \right],
\en\ee
in terms of 8 real components of $\z^0 = (\z^0_1,\ldots,\z^0_8)$. Taking $c_1=1, c_2=0$ in~\eqref{grav-sol} and setting $\z^0_a = \d_{ab}$ for some $b \in \{1,\ldots,8\}$ we get eight basis Killing spinors $\e_b$. There are eight more Killing spinors arising when $c_1=0, c_2=1$. Explicit computation then shows, that taking other values of the parameters $c_1, c_2$ in~\eqref{chi-otvet} results in the same set of Killing spinors, up to relabeling.

Note that as in the $AdS_7 \times M_3$ case, the 16 basis Killing spinors can be labeled in such a way that $\e_1,\ldots,\e_8$ are the Poincar\'e Killing spinors, which only depend on the radial $AdS_5$ coordinate $\r$, while $\e_9,\ldots,\e_{16}$ are the superconformal Killing spinors that depend on all $AdS_7$ coordinates.

\section{Fermionic T-duality}
\label{ferm}

The Killing spinors that we have found can be used to study fermionic T-duals of the supergravity solutions of the sections~\ref{ads7},~\ref{ads5}. We will proceed to this after briefly reviewing the rules that link fermionic T-dual backgrounds (fermionic Buscher rules). For more detail on the basics of fermionic T-duality see the reviews~\cite{OColgain:2012si,Bakhmatov:2011ab} or the original derivation~\cite{Berkovits:2008ic}. 

Fermionic T-duality only transforms the RR fluxes and the dilaton $\f$; there is no change in the metric nor in the antisymmetric $b$ field. In type IIA supergravity it is convenient to unify the RR field strengths $F_{\m\n}$ and $F_{\m\n\r\s}$ together with the Romans mass parameter $m$ into a bispinor ${F^\a}_\b$ (for the spinor and gamma matrix conventions see appendix~\ref{gam}):
\be
\label{gamma-expansion}
{F^\a}_\b = m\, \d^\a_\b + \frac{1}{2!} F_{\m\n} {(\g^{\m\n})^\a}_\b + \frac{1}{4!} F_{\m\n\r\s} {(\g^{\m\n\r\s})^\a}_\b.
\ee
Fermionic T-duality transformation rules are:
\be
\label{dilaton}
\begin{gathered}
e^{2\f'} = e^{2\f}\,\det C, \\
{F'^\a}_\b = (\det C)^{-1/2} \left( {F^\a}_\b + 16 i e^{-\f}\, C_{IJ}^{-1}\, \e_I^\a \e_{J\b} \right),
\end{gathered}
\ee
where $C_{IJ}$ is the matrix defined by
\be
\label{C}
\p_\m C_{IJ} = i\, \overline{\e}_I \G_\m \G^{11} \e_J.
\ee
The transformation parameters $\e_I$ are the Killing spinors of the original background. Indices $I,J$ run over the subset of the Killing spinors that we have chosen to T-dualise. In particular, one may choose to do fermionic T-duality with respect to just one Killing spinor, in which case the $I,J$ indices become redundant and $C_{IJ}$ is no longer a matrix but just some scalar function $C$. For consistency of the above transformation, the Killing spinors must obey the so called abelian constraint 
\be\label{abelian}
\overline{\e}_I \G_\m \e_J \overset{!}{=} 0,
\ee
which comes from the requirement that the corresponding supersymmetries anticommute~\cite{Berkovits:2008ic}. Alternatively, the abelian constraint may be interpreted as integrability condition for~\eqref{C}~\cite{Godazgar:2010ph}. 

Note that~\eqref{abelian} is a nontrivial constraint even for a single Killing spinor, i.e. when $I=J$. One can check by explicit computation that none of the basis Killing spinors $\e_a$ found in the sections~\ref{killing-7x3},~\ref{killing-5x5} satisfy the abelian constraint. In fact, this constraint can never be satisfied by a Majorana spinor, which makes it necessary to complexify the Killing spinors. Following the traditional approach we break the Majorana condition by considering instead of the basis elements $\e_a$ their complex linear combinations of the form $\e = \e_a + i\e_b$. For some choices of $a$ and $b$ the abelian constraint can be satisfied by $\e$. Then the equations~\eqref{dilaton},~\eqref{C} give a fermionic T-dual background that solves the supergravity field equations and is guaranteed to preserve the same amount of supersymmetry as the original solution. However, the dual background in general is not real as a consequence of the complexification of the Killing spinors. 

In both the $AdS_7 \times M_3$ and the $AdS_5 \times M_5$ cases there are sixteen basis Majorana Killing spinors $\e_a$. For a generic complex linear combination $\e = \e_a + i\e_b$ (assuming that $a \neq b$) the abelian constraint takes the form:
\be\label{abelian1}
(\overline\e_a + i\,\overline\e_b)\, \G^\m (\e_a + i\e_b) = \overline\e_a \G^\m \e_a - \overline\e_b \G^\m \e_b + 2i\, \overline\e_a \G^\m \e_b \overset{!}{=} 0.
\ee

Let us first confine our attention to the Poincar\'e Killing spinors $\e_1,\ldots,\e_8$, which have a simpler algebraic structure and can be treated generically. Recall that these are the Killing spinors that result from keeping the $\z_-^0$ part of the complete $AdS$ Killing spinor~\eqref{zeta}, \eqref{zeta1}. For the $AdS_7 \times M_3$ Killing spinors of the section~\ref{killing-7x3}, explicit computation shows that the vectors $v^\m_a = -\frac{1}{16}\overline{\e}_a \G^\m \e_a$ are lightlike, pointing in the negative $x^5$ direction for $a \in \{1,2,3,4\}$, and pointing in the positive $x^5$ direction for $a \in \{5,6,7,8\}$:
\be\label{v}
\bn
v^\m_a &= (1,0,\ldots,0,-1,0,\ldots,0), & a \in \{1,2,3,4\},\\
v^\m_a &= (1,0,\ldots,0,1,0,\ldots,0), & a \in \{5,6,7,8\}.
\en
\ee
Thus the first two terms in~\eqref{abelian1} cancel each other if and only if either both $a, b \in \{1,2,3,4\}$, or both $a, b \in \{5,6,7,8\}$. As to the last term in~\eqref{abelian1}, the vector $u^\m_{ab} = \overline{\e}_a \G^\m \e_b$ with $a\neq b$ also vanishes if and only if either both $a, b \in \{1,2,3,4\}$, or both $a, b \in \{5,6,7,8\}$. To summarise, a complexified Killing spinor
\be
\e = \e_a + i \e_b, \qquad a\neq b
\ee
is a valid parameter for a fermionic T-duality transformation of~\eqref{dilaton},~\eqref{C}, whenever $a,b \in \{1,2,3,4\}$ or $a,b \in \{5,6,7,8\}$.

\subsection{Fermionic T-duals of $\7$}

As an example of a dual background, consider fermionic T-duality generated by the Killing spinor $\e = \e_1 + i \e_2$ of the $AdS_7 \times M_3$ background of the section~\ref{ads7}. The equation~\eqref{C} takes the form $\partial_\m C = i\, \overline\e\, \G_\m \G^{11} \e$, which vanishes as can be checked by direct computation. Thus $C$ in this case is an arbitrary constant, which can be set to one in order to keep the value of the dilaton fixed~\eqref{dilaton}. Then the original fluxes~\eqref{flux-7x3} and the mass parameter $m$ also do not change, however the last term in the second of the equations~\eqref{dilaton} creates new components of the RR flux. We find the following 32 components of the RR 4-form:
\be\label{dual-flux}
\bn
F_{abc6} &= -K \cos\b,\\
F_{abc7} &= i K,\\
F_{abc8} &= K \sqrt{1-x^2} \sin\b,\\
F_{abc9} &= K x \sin\b,
\en
\ee
where the indices $abc$ can take the values $013$, $014$, $023$, $024$, $135$, $145$, $235$, $245$. The values of $K$ and $abc$ are related in the following way:
\be\bn
abc &=  013, 135 \qquad  & K = 16\, i \r\, e^{A-\f-i\th},\\
abc &=  024, 245 \qquad  & K = -16\, i \r\, e^{A-\f-i\th},\\
abc &=  014, 145, 023, 235 \qquad  & K = 16 \r\, e^{A-\f-i\th}.
\en\ee
There are no new components of the RR 2-form. Up to the above 32 new components, the solution remains the same as described in the section~\ref{killing-7x3}.

This is a characteristic form of a fermionic T-dual solution whenever the complexified Killing spinor is constructed as explained above. Apart from $\e = \e_1 + i \e_2$ one may consider, e.g. $\e_3 + i\e_4$, $\e_5 + i\e_6$, and $\e_7 + i\e_8$. Each of these gives a fermionic T-dual RR flux same as above, but the values of the indices $abc$ are slightly different every time. The resulting fermionic T-dual may be simplified considerably if these four Killing spinors are dualised at the same time. To achieve this, take the Killing spinor $\e_I$ that appears in~\eqref{dilaton},~\eqref{C} to assume the values just listed,
\be\label{4fTd}
\e_I = 
	\left\{
		\bn
			\e_1 + i\e_2, \qquad & I = 1,\\
			\e_3 + i\e_4, \qquad & I = 2,\\
			\e_5 + i\e_6, \qquad & I = 3,\\
			\e_7 + i\e_8, \qquad & I = 4.\\
		\en
	\right.
\ee
The abelian constraint~\eqref{abelian} is satisfied for any pair $I$ and $J$, and the derivatives of $C_{IJ}$~\eqref{C} are vanishing as before. We can take $C_{IJ}$ to be a unit matrix, in which case the dilaton and the original fluxes keep their values, whereas almost all the contributions to the RR 4-form coming from different $\e_I$ cancel each other. There are only 8 new components of the 4-form surviving, which are given by the same expressions~\eqref{dual-flux} with $abc = 023$, $145$ and an extra factor of four.

The argument after the equation~\eqref{v} allows for multiple other complexification patterns. We note here one more specific case that leads to a simple fermionic T-dual, and later we will draw certain conclusions valid for any choice of complexification. Consider the following Killing spinors:
\be\label{v1}
\e_I = 
\left\{
\bn
\e_1 + i\e_3, \qquad & I = 1,\\
\e_2 + i\e_4, \qquad & I = 2,\\
\e_5 + i\e_7, \qquad & I = 3,\\
\e_6 + i\e_8, \qquad & I = 4.\\
\en
\right.
\ee
Again we are free to choose $C_{IJ}$ to be a unit matrix, hence the background is the same as before, up to 12 new RR 4-form components:
\be
\bn
F_{abc6} &= K \sin\b,\\
F_{abc8} &= K \sqrt{1-x^2} \cos\b,\\
F_{abc9} &= K x \cos\b.\\
\en
\ee
The values of $K$ and $abc$ are now given by:
\be
\bn
abc = 123, \qquad & K = 64\, i \r\, e^{A-\f};\\
abc = 124, \qquad & K = -64 \r\, e^{A-\f};\\
abc = 356, \qquad & K = 64 \r\, e^{A-\f};\\
abc = 456, \qquad & K = 64\, i \r\, e^{A-\f}.\\
\en
\ee
In other words, choosing the complexification scheme~\eqref{v1} has allowed us to eliminate any $\th$ dependence in the fermionic T-dual background. However, as in the previous examples, the dual solution is not real.

\subsection{Fermionic T-duals of $\5$}

Turning now to the case of the $\5$ background of section~\ref{ads5}, we encounter similar patterns of the RR fluxes in the fermionic T-dual backgrounds. However, the results here are more cumbersome in comparison to the $\7$ results, as are the Killing spinors themselves. When fermionic T-dualising one Poincar\'e supersymmetry in $\5$ we get 32 RR flux components, the same amount as in $\7$. However, the form of these fluxes is more intricate. Certain simplification can be observed in the case of a combination of four Poincar\'e supersymmetries, similar to~\eqref{4fTd}. In this case fermionic T-duality creates 16 new RR flux components, including the 2-form: 
\be\label{dual}
\bn
F_{57} &= K \sin(\c+\ps) \sin\th \sin\varphi,\\
F_{58} &= K \sin(\c+\ps) \cos\th,\\
F_{59} &= K \cos(\c+\ps),\\
F_{67} &= K \cos(\c+\ps) \sin\th \sin\varphi,\\
F_{68} &= K \cos(\c+\ps) \cos\th,\\
F_{69} &= -K \sin(\c+\ps),\\
\en
\ee
where $K=-128 \r e^{A-\f}$. Note that we are able to obtain a manifestly real RR 2-form flux. However, an imaginary 4-form flux appears, $F_{23ab} = i F_{ab}$, where $ab = 57,58,59,67,68,69$, and finally 4 more components of the 4-form emerge, 
\be
F_{0345} = -F_{0346} = -i F_{5789} = -i F_{6789} = -i K \cos(\c+\ps) \sin\th \cos\varphi.
\ee

Note that up to know we have exclusively used Poincar\'e Killing spinors as an input for fermionic T-duality. Both in the $\7$ and in the $\5$ case it is possible to fermionic T-dualise the background with respect to a superconformal Killing spinor as well. These are the Killing spinors whose $AdS$ part has dependence on the flat $AdS$ coordinates, which result from keeping the $\z_+^0$ part of the complete $AdS$ Killing spinor~\eqref{zeta}, \eqref{zeta1}. The resulting fermionic T-dual RR fluxes are rather intricate and we will not give their explicit form here. These expressions are akin to what was classed as the `complicated' fermionic T-dual case in \cite{Bakhmatov:2009be}, or as a T-dual with respect to the supernumeracy Killing spinors in \cite{Bakhmatov:2011aa}. They are similar to the fluxes of~\eqref{dual}, additionally multiplied by a polynomial of degree up to 4 in the $AdS$ coordinates. No matter what kind of a Killing spinor we use, the fermionic T-duality parameter $C$ appears always to be a constant.

\subsection{Constant fermionic T-duality parameter}
\label{const}

All the fermionic T-dual backgrounds described above have a property that the duality parameter $C$ is a constant, $\p_\m C = 0$. Vanishing of the corresponding Killing spinor contraction~\eqref{C} is by no means obvious and must be checked by direct computation. However, in the simpler case of the $AdS_7 \times M_3$ solution it is possible to prove that $\p_\m C = 0$ for an arbitrary choice of the Killing spinors. 

Consider two arbitrary supersymmetries $\e_I = \sum_{a=1}^{16} k_a \e_a$ and $\e_J = \sum_{a=1}^{16} l_a \e_a$ subject to the condition $\overline\e_I \G^\m \e_J = 0$. Note that we now consider all possible Killing spinors, Poincar\'e as well as superconformal. Plugging in the values of the $AdS_7 \times M_3$ Killing spinors we arrive at the following equations for $\m = 7,8,9$:
\be\label{comb}
\begin{gathered}
k_a \, (i\s_2 \otimes \g^0)_{ab} \, l_b = 0,\\
k_a \, (\s_1 \otimes \mathbb{1})_{ab} \, l_b = 0,\\
k_a \, (\s_3 \otimes \mathbb{1})_{ab} \, l_b = 0.
\end{gathered}
\ee
There is no need to consider remaining values of $\m$ as the above three equations fully constrain the parameter of fermionic T-duality. It turns out that $\p_\m C_{IJ} = i \overline\e_I \G_\m \G^{11} \e_J$ for any $\m \in \{0\ldots,9\}$ can be completely expressed in terms of the polynomials in~\eqref{comb}. Hence the abelian condition $\overline\e_I \G^\m \e_J = 0$ implies the vanishing of $\p_\m C_{IJ}$ and we are free to choose constant values of $C_{IJ}$ for any pair of supersymmetries in the theory. The effect of fermionic T-duality~\eqref{dilaton} is then to rescale the string coupling $e^\f$ and the RR flux ${F^\a}_\b$ by powers of the constant $\det C_{IJ}$. What makes the transformation nontrivial is the additive correction to the fluxes coming from the last term in~\eqref{dilaton}. This term leads to the new components of the RR flux that we have seen in the examples above.

Turning now to the $AdS_5 \times M_5$ case we find that the Killing spinors are more complicated and do not lend themselves to an analogous treatment. Nevertheless, all particular abelian combinations of the Killing spinors that we have tried give $\p_\m C_{IJ} = 0$, which suggests that $C_{IJ}$ is always a constant for the $AdS_5 \times M_5$ backgrounds, similarly to the $\7$ case.

\subsection{Constant Romans mass parameter}

The behaviour of IIA mass parameter $m$ under fermionic T-duality is a long standing question~\cite{OColgain:2012si} which was one of the motivations for this work. Recall that we have incorporated $m$ into the bispinor of RR fields~\eqref{gamma-expansion}, which under fermionic T-duality has the transformation law:
\be
{F'^\a}_\b = (\det C)^{-1/2} \left({F^\a}_\b + 16 i e^{-\f} {A^\a}_\b\right),
\ee
where ${A^\a}_\b = C_{IJ}^{-1}\, \e_I^\a \e_{J\b}$. Taking the trace of this relation we remove the 2-form and the 4-form terms in the gamma-matrix expansion of $F$ and $F'$. This leaves us with the transformation law of the mass parameter under fermionic T-duality:
\be
\label{mass}
m' = (\det C)^{-1/2} \left( m + i e^{-\f} \,\mathrm{tr}\, A \right).
\ee
Thus, $m$ is shifted by the trace part of the Killing spinor matrix $A$. In particular, this shift might in principle generate mass in some type IIA background that was originally massless. The rescaling by a factor of $(\det C)^{-1/2}$ in the present case is trivial as $C_{IJ}=\mathrm{const}$, but in general $C_{IJ}$ can be a coordinate dependent function. Note that the Romans mass parameter is intrinsically a constant quantity, hence a nontrivial coordinate dependent $C_{IJ}$ would require a very special form of $\mathrm{tr}\, A$, so as to keep both $m$ and $m'$ constants.

For the $AdS_7 \times M_3$ and $AdS_5 \times M_5$ solutions of the sections~\ref{ads7},~\ref{ads5}, however, the Killing vectors result in a matrix $A$ that is traceless. Similarly to the previous subsection, this can be proved strictly for the $AdS_7 \times M_3$ case, and is very likely to hold in general in $AdS_5 \times M_5$ as well.

Consider again two arbitrary anticommuting Killing spinors, i.e. two linear combinations $\e_I = \sum_{a=1}^{16} k_a \e_a$ and $\e_J = \sum_{b=1}^{16} l_b \e_b$ with coefficients subject to~\eqref{comb}. Assume that we choose some nonzero constant value for the parameter $C_{IJ}$. Explicit computation then shows that the trace of the matrix ${A^\a}_\b = C_{IJ}^{-1}\, \e_I^\a \e_{J\b}$ is expressed in terms of the same polynomials that appear in the constraints~\eqref{comb}. Hence, the anticommutation constraint for the supersymmetries implies that the only possible transformation of the mass parameter in the present case is rescaling by a constant $(\det C)^{-1/2}$. 

\section{Discussion}
\label{conc}

In this article the Killing spinors of the $AdS_7 \times M_3$ and $AdS_5 \times M_5$ backgrounds of massive type IIA supergravity were derived. We have studied various fermionic T-dual backgrounds that are parameterised by the Killing spinors. In general, the dual background can be characterised by a constant rescaling of the Romans mass parameter $m$, the string coupling $e^\f$, and the RR fluxes ${F^\a}_\b$. There are some new components of the RR flux as well. At the same time, the geometry and the NSNS 2-form field $b$ are fixed to their original values, which is a generic property of fermionic T-duality. Essentially we have presented a way of deforming the original backgrounds by introducing extra RR field components while keeping all the supersymmetries.

A typical fermionic T-dual RR flux is given in~\eqref{dual-flux}. It is interesting to note that although adding up the contributions from different commuting Killing spinors reduces the number of new RR flux components, it is impossible to eliminate all the new contributions and just keep the original fluxes. Even the maximum commuting subset of complexified Killing spinors considered in~\eqref{4fTd}, while canceling almost all the fluxes that result from individual Killing spinors, still does not lead to exact self-duality. In fact, all known self-T-duality setups in the $AdS_m \times M_n$ spacetimes~\cite{Berkovits:2008ic,OColgain:2012ca,Abbott:2015mla,Abbott:2015ava} also required performing bosonic T-dualities along the $k-1$ flat directions of $AdS_k$, including a timelike T-duality. This transformation generates a characteristic contribution of $\d_B\f = (k-1) \log \r$ to the dilaton, where $\r$ is the $AdS$ radial coordinate. In the self-dual cases this contribution can be canceled by fermionic T-duality, which adds to the dilaton an extra term $\d_F \f = \frac12 \log \det C = - \d_B \f$. However, in the section~\ref{const} we have shown that the parameter $C$ for the solutions of~\cite{Apruzzi:2013yva,Apruzzi:2015zna} can only assume constant values, hence the standard self-T-duality scheme does not work here. This agrees with the classification of self-T-dual backgrounds constructed in~\cite{Adam:2009kt,Dekel:2011qw}.

We have seen in the section~\ref{ferm} that the RR 2-form flux rarely appears in the fermionic T-dual. In fact, for the $\7$ solutions none of the Killing spinors produce any contributions to the RR 2-form. On the contrary, in $\5$ the 2-form flux routinely appears after fermionic T-duality and we have seen the example of it in~\eqref{dual}.

The new components of the RR 4-form that appear after fermionic T-duality in the $\7$ background could not be found in the original study of~\cite{Apruzzi:2013yva} because the Ansatz of the $AdS_7 \times M_3$ solution that was used there intentionally did not include any 4-form flux. This restriction was put in order to protect the $AdS_7$ symmetry, because a nonvanishing 4-form necessarily would a have at least one leg off the internal $M_3$ manifold. The fermionic T-dual solutions with the 4-form flux found here nevertheless keep the same $AdS_7$ geometry. This is possible essentially because the fermionic T-dual fluxes that we have found do not backreact: they have vanishing energy-momentum tensor and therefore do not contribute to the gravity field equations. Vanishing stress-energy of some fermionic T-dual fluxes is a feature already observed in~\cite{Bakhmatov:2009be} for the fermionic T-duals of D-branes. Note also that the D-brane dimension is not modified by fermionic T-duality~\cite{Berkovits:2008ic}, thus the fact that there are new RR fluxes does not imply that there are new D-brane sources. In fact, it is possible that any fermionic T-dual RR flux is a solution of the field equations without sources. In that case the new fluxes are decoupled from the rest of the fields in the solution and they naturally do not break any symmetries.

\section*{Acknowledgements}

The author would like to thank Edvard Musaev and Alessandro Tomasiello for stimulating discussions and helpful comments. This work was supported by the RFBR grant 14-02-31494 and by the Russian Government program of competitive growth of Kazan Federal University.

\appendix

\section{Index summary}

Various indices that we use are:
\begin{align*}
&& \m,\n &\in \{0, \ldots, 9\} && \mathrm{Lorentz~vector~index},&&\\
&& \a,\b &\in \{1, \ldots, 16\} && \mathrm{Weyl~spinor~index},&&\\
&& a,b &\in \{1, \ldots, 16\} && \mathrm{counts~linearly~independent~Killing~spinors},&&\\
&& I,J &\in \{1, \ldots, X\} && \mathrm{for~some~} X\leq16: \mathrm{subset~of~the~Killing~spinors}&&\\
&&     &               && \mathrm{to~be~dualised}.&&
\end{align*}
The indices used for splitting the $10d$ vector index $\mu$ into $7d$ and $3d$ vectors are
\begin{align*}
m,n &\in \{0,\ldots,6\}, \\
i,j &\in \{1,2,3\},
\end{align*}
or, for the $5+5$ split:
\begin{align*}
m,n &\in \{0,\ldots,4\}, \\
i,j &\in \{1,\ldots,5\} 
\end{align*}

\section{Spinor conventions}
\label{gam}

Both the $AdS_5 \times M_5$ and the $AdS_7 \times M_3$ gamma-matrices defined below are of the block form
\be
\begin{aligned}
\G^\m =
    \left(
        \begin{array}{cc}
            0 & \g^{\m\,\a\b}\\
            \g^\m_{\a\b} & 0\\
        \end{array}
    \right),
\end{aligned}
\qquad
\m\in \{0,\ldots,9\}, 
\ee
where $\a,\b \in \{1,\ldots,16\}.$ The $16 \times 16$ matrices $\g^{\m\,\a\b}$ and $\g^\m_{\a\b}$ appear in the decomposition of the RR flux bispinor~\eqref{gamma-expansion}. This spinor index convention means that 16-component left and right Weyl spinors have indices $\e^\a$ and $\e_\a$:
\be
\e = \left(
        \begin{array}{c}
            \e^\a \\
            \e_\a\\
        \end{array}
    \right).
\ee
When a spinor $\e$ appears without an index, it means the full 32 component spinor as above. Using the explicit gamma-matrix realizations below, it is easy to verify that we are dealing with Weyl representation as the $d=10$ chirality operator is in its conventional form $\G^{11} = \G^0 \ldots \G^9 = \mathbb{1}_{16} \otimes \s^3$.

In order to find the Killing spinors of a supergravity background we require that the supersymmetry variations of all fermionic fields in the theory vanish. Variations of the massive type IIA fermions in the conventions of~\cite{Bergshoeff:2001pv} are given by
\be\label{variations}\bn
\d\l &= \left( \slashed\p\f +\frac12 \slashed H \,\G^{11} \right)\e + \frac{e^\f}{4} \left( 5m + 3 \slashed F_{(2)} \G^{11} + \slashed F_{(4)} \right)\e,\\
\d\ps_{\m} &= \left( D_{\m} + \frac14 \G^{11} \slashed H_{\m} \right)\e + \frac{e^\f}{8} \left( m \G_{\m} + \slashed F_{(2)} \G_{\m} \G^{11}  + \slashed F_{(4)} \G_{\m}\right)\e,
\en\ee
where $D_{\m} = \p_{\m} + \frac12 \slashed \w_{\m}$ and the $1/n!$ factors have been absorbed in the definition of slash, $\slashed F_{(n)} = \frac{1}{n!} F_{\m_1\ldots\m_n} \G^{\m_1\ldots\m_n}$.

For the sake of defining Majorana spinors, we consider the standard intertwiners $B,C,D$, which map gamma-matrices to their complex conjugate, transpose, and hermitian conjugate, respectively. Majorana condition is equivalent to requiring that Dirac and Majorana conjugations of a spinor $\e$ agree, $\e^\dagger D = \e^T C$. Using the definitions of $B,C,D$ below, this implies
\be
\e = -B\e^*.
\ee
We use Majorana conjugation to build Lorentz tensors from spinors, e.g. $\overline\e_1 \G^\m\e_2 = \e_1^T C \G^\m\e_2$.

Any type IIA fermion $\e$ should be decomposable into two Majorana-Weyl spinors $\e_{1,2}$ of opposite chiralities:
\be
\e_{1,2} \overset{!}{=} -B \e_{1,2}^*,\qquad \e_1 \overset{!}{=} \G^{11} \e_1,\qquad \e_2 \overset{!}{=} -\G^{11} \e_2.
\ee
These Majorana-Weyl spinors may be further decomposed into the $AdS$ part $\z$ and the $M$ part $\c$:
\be\label{susy}\bn
\e_1 &= (\z \otimes \c_1 + \z^c \otimes \c_1^c) \otimes v_+,\\
\e_2 &= (\z \otimes \c_2 - \z^c \otimes \c_2^c) \otimes v_-.
\en\ee
Note that since in both gamma-matrix representations below $\G^{11} = \mathbb{1}\otimes\s^3$, the chirality constraints immediately imply that $\s^3 v_{\pm} = \pm v_{\pm}$, hence $v_+ = {1 \choose 0}$, $v_- = {0 \choose 1}$. Charge conjugation for the component spinors $\z^c, \c^c$ is defined according to the decomposition of the charge conjugation matrix $B$.

Occasionally it proves more convenient to work with 16-component spinors $\varepsilon_{1,2}$ defined as
\be\label{epsilon-5x5}
\e = \ve_1\otimes v_+ + \ve_2\otimes v_- = {\ve_1 \choose \ve_2} = {\z \otimes \c_1 + \z^c \otimes \c_1^c \choose \z \otimes \c_2 - \z^c \otimes \c_2^c}.
\ee

\subsection{Gamma-matrices for $d=7+3$ spacetime}

In choosing the representation we mostly follow the conventions outlined in~\cite{Apruzzi:2013yva}. The decomposition appropriate for this case is in terms of the Lorentzian $d=7$ gamma-matrices $\a^m$ and the Euclidean $d=3$ gamma-matrices $\s^i$ (Pauli matrices):
\be\label{gamma7x3}\bn
\G^m = \a^m \otimes \mathbb{1}_2 \otimes \s^2&,\qquad m\in \{ 0,\ldots, 6 \},\\
\G^{i+6} = \mathbb{1}_8 \otimes \s^i \otimes \s^1&,\qquad i\in \{ 1,2,3 \}.
\en\ee
We can choose the $d=7$ matrices to be
\be\label{small-gamma7x3}\bn
\a^0 &= \mathbb{1}_2 \otimes i\s^2 \otimes \s^1,\\
\a^1 &= \mathbb{1}_2 \otimes  \s^2 \otimes \s^3,\\
\a^2 &= \s^1 \otimes  \mathbb{1}_2 \otimes \s^2,\\
\a^3 &= \s^3 \otimes  \mathbb{1}_2 \otimes \s^2,\\
\a^4 &= \s^2 \otimes \s^1 \otimes  \mathbb{1}_2,\\
\a^5 &= \s^2 \otimes \s^3 \otimes  \mathbb{1}_2,\\
\a^6 &=-\s^2 \otimes \s^2 \otimes  \s^2.\\
\en\ee
This choice gives rise to the following intertwiners:
\be\bn
B &= \a^0 \otimes \s^2 \otimes \s^3,\\
C &= i\G^8,\\
D &= \G^0,
\en\ee
which satisfy
\be\bn
B \G^\m B^{-1} &= +(\G^\m)^*,\\
C \G^\m C^{-1} &= -(\G^\m)^T,\\
D \G^\m D^{-1} &= -(\G^\m)^\dagger.
\en\ee

From the decomposition of $B$ it follows that charge conjugation for component spinors is defined by $\z^c = -\a^0\z^*$, $\c^c = \s^2 \c^*$.

\subsection{Gamma-matrices for $d=5+5$ spacetime}

Gamma-matrices used in the section~\ref{ads5} are decomposed into the $AdS_5$ matrices $\a^m$, and the $M_5$ matrices $\b^i$:
\be\label{gamma5x5}\bn
\G^m = \a^m \otimes \mathbb{1}_4 \otimes \s^2&,\qquad m\in \{ 0,\ldots, 4 \},\\
\G^{i+4} = \mathbb{1}_4 \otimes \b^i \otimes \s^1&,\qquad i\in \{ 1,\ldots,5 \}.
\en\ee
We choose the $d=5$ Lorentzian gamma-matrices in the form
\be\label{small-gamma5x5}\bn
\a^0 &=i\s^2 \otimes \s^1,\\
\a^1 &= \s^2 \otimes \s^3,\\
\a^2 &= \s^1 \otimes \mathbb{1}_2,\\
\a^3 &= \s^3 \otimes \mathbb{1}_2,\\
\a^4 &= \s^2 \otimes \s^2.
\en\ee
The $d=5$ Euclidean gamma-matrices are $\b^1 = -i\a^0$ and $\b^{2,3,4,5}=\a^{1,2,3,4}$. We have the following intertwining operators:
\be\bn
B &= \a^1 \otimes i\b^1\b^2 \otimes \s^3,\\
C &= \a^0\a^1 \otimes \b^1\b^2 \otimes \s^1,\\
D &= \G^0,
\en\ee
satisfying:
\be\bn
B \G^\m B^{-1} &= -(\G^\m)^*,\\
C \G^\m C^{-1} &= +(\G^\m)^T,\\
D \G^\m D^{-1} &= -(\G^\m)^\dagger.
\en\ee
Note that $B^* = B = B^{-1}$. Charge conjugation of the component spinors in this case is given by $\z^c = -\a^1\z^*$, $\c^c = i\b^1\b^2 \c^*$.

\section{$AdS$ Killing spinors}
\label{ads}

The $AdS$ part of the Killing spinor $\z$, which appears in~\eqref{susy} and~\eqref{epsilon-5x5}, may be computed straightforwardly from the condition of zero supersymmetry variation of the gravitino with an $AdS$ index $\d\ps_m = 0$. For example, in the $AdS_7 \times M_3$ case, using the background fields of section~\ref{ads7} in the variation~\eqref{variations}, we get
\be
\d\ps_m = D_m {\ve_1 \choose \ve_2} + \frac{i m e^\f}{8} {-(\a_m \otimes \mathbb{1}) \ve_2 \choose (\a_m \otimes \mathbb{1}) \ve_1 } - \frac{e^\f}{8}\, F_{89} {(\a_m \otimes \s^1) \ve_2 \choose (\a_m \otimes \s^1) \ve_1}.
\ee
Using the result~\eqref{dil-otvet} for the internal spinors $\c_{1,2}$, this may be simplified to
\be
\d\ps_{m} = {\d_{m}\z\otimes\c_1 + \d_{m}\z^c\otimes\c_1^c \choose \d_{m}\z\otimes\c_2 - \d_{m}\z^c\otimes\c_2^c },
\ee
where
\begin{align}
&&\d_{\u m}\z &= \p_{\u m}\z + \r\, \a_m\, \frac{1+\a_6}{2}\,\z,  &&\mathrm{when}\: m \in \{0,\ldots,5\},&&\\
&&\d_{\u \r} \z &= \p_{\u\r}\z + \frac{1}{2\r} \a_6 \z,  &&\mathrm{when}\: m=6.&&
\end{align}
$\d_{\u m}\z = 0$ are the usual $AdS$ Killing spinor equations~\cite{Lu:1996rhb}. These can be solved easily using the fact that any solution $\z$ to these equations can be decomposed into the sum of eigenvectors of $AdS$ radial gamma-matrix $\g_6$,
%\footnote{When comparing to~\cite{Lu:1996rhb}, one should make replacement $\g_i \to -\g_i$ and $\z_\pm^0 \to \z_\mp^0$ in the equations and in the solution given here. Then one gets the equations and the solution of~\cite{Lu:1996rhb} up to a sign error in their paper.}
\be
\z = \z_+ + \z_-,\qquad \z_\pm = \frac{1\pm\a_6}{2}\,\z,\qquad \a_6\z_\pm = \pm\z_\pm.
\ee
The solution is
\be\label{ads-spinor}
\z = \r^{1/2} \z_-^0 + \left( \r^{-1/2} - \r^{1/2} x^m \a_m \right) \z_+^0, \qquad m\in\{0,\ldots,5\},
\ee
which is parameterized by an arbitrary constant spinor $\z^0$. For the case of $AdS_7$ this has 8 components. In the $AdS_5 \times M_5$ case the above derivation is entirely analogous with the same result, where now $\z^0$ is a 4-component spinor. %Together with the two complex parameters $c_{1,2}$ of the internal spinors~\eqref{chi-otvet}, we get a total of 16 complex or 32 real parameters. 

\bibliographystyle{utphys}
\bibliography{bib}

\end{document}